# Metal Mesh-Based Infrared Transparent Electromagnetic Shielding Windows with Balanced Shielding Properties over a Wide Frequency Range

**Yuanlong Liang, Kui Wen, Zhaofeng Wu, Jisheng Pan, Wencong Liu, Lixiang Yao, Peiguo Liu and Xianjun Huang***

College of Electronic Science and Technology, National University of Defense Technology, Changsha, 410073, China

* Correspondence: huangxianjun@nudt.edu.cn (X.H.)

**Abstract**

With the increasing complexity of the electromagnetic environment, electromagnetic interference has already been an important problem for optoelectronic systems to be reckoned with. The metal mesh film is a kind of widely used electromagnetic shielding material with both visible and infrared transparency. However, the shielding performance of previously reported mesh materials is frequency dependent. Here, we report an infrared-transparent electromagnetic shielding windows based on metal mesh with irregular patterns. The mesh coatings are prepared on sapphire substrate using ultraviolet photolithography technology, and provide efficient electromagnetic shielding effectiveness of more than 20 dB in the wide frequency range of 1.7–18 GHz while maintaining high infrared optical transparency. In addition, there is no distinct variation in shielding effectiveness between low and high frequency range, exhibiting a balanced shielding characteristic throughout a broad frequency band. This work could be significant in protecting infrared optoelectronic devices from electromagnetic interference.

# 1. Introduction

With the rapid development of electronic and wireless communications, electromagnetic radiation generated by the operation of various electronic and electrical equipment and related infrastructure has been widely present in the space. Meanwhile, the ensuing electromagnet radiation are broadening in frequency spectrum and increasing in intensity, resulting in an increasingly complex electromagnetic environment [1,2]. Electromagnetic interference (EMI) will be caused by disorderly, high-intensity electromagnetic radiation, which can disrupt the normal operation of electronic information systems [3]. Especially for an optoelectronic system, its internal sophisticated optical sensors, photoelectric conversion modules, and integrated circuits are particularly susceptible to electromagnetic radiation interference [4,5]. Given that the optical window is the primary channel via which electromagnetic radiation enters the optoelectronic system, electromagnetic protection of optoelectronic systems can be performed by incorporating electromagnetic shielding into the optical windows.

Transparent conductive films can be applied as EMI shielding materials of optical widows or domes for the suppression of undesirable electromagnetic radiation. Currently, researches on transparent EMI shielding materials are mostly focused on visible transparent materials, however, infrared optoelectronic systems play a significant role in the military, aerospace, and other fields. Therefore, with the actual needs, it is more urgent to develop infrared transparent electromagnetic shielding windows. Common strategies for achieving infrared transparent electromagnetic shielding include applying ultra-thin metal film or metal mesh film to the outer surface of the window or utilizing infrared transparent conductive oxides [6-10]. Besides, some new materials, such as carbon nanomaterials, also present favorable conductive and infrared optical properties [11,12].

EMI shielding materials for optical windows of optoelectronic systems not only require efficient EMI SE to block microwave radiation, but also need to be highly transparent for high-quality detecting and imaging [13,14]. Continuous or quasi-continuous transparent conducting films, such as transparent conductive oxides, ultra-thin metals, and films of carbon nanotubes can provide efficient EMI shielding effectiveness (SE) [15-17]. However, a well-known problem with these transparent conductive films is that the transparency is inevitably affected by the increase of loading amount or film thickness[18,19]. It follows that decoupling the optical transmittance and EMI SE is challenging. The perforated metallic micro-structures or networks, like metallic meshes, can alleviate the restraint between optical transmittance and electrical conductivity to some extent by increasing mesh thickness [20]. It has been reported that metallic meshes with micron/submicron line width and submillimeter grid spacing allow for the high transmission of Vis-IR light and strong reflection of microwaves simultaneously [21,22]. For this type of material, the void aperture contributes to the optical transparency while interconnected fine metallic wires supply the film with electrical conductivity. While the overriding concern is with EMI shielding effectiveness, the shielding frequency spectrum is a potential problem that cannot be ignored. Mesh-based shielding materials usually present frequency-dependent shielding effectiveness, with shielding effectiveness decreasing as frequency increases [21,23,24]. Therefore, the shielding capability of the metal mesh in a broad frequency is mainly measured by the shielding performance at high frequencies. As mesh thickness increases, however, the enhancement in shielding effectiveness varies between bands of high and low frequencies. Highly efficient mesh-based EMI shielding materials have still been underachieving. It is expected that outstanding overall performance can be realized by engineering and optimizing the structural parameters, i.e., materials, density, mesh pattern, linewidth, spacing and, particularly the film thickness.

Herein, we demonstrated an IR transparent EMI shielding window with efficient and stable EMI shielding properties over broadband frequency spectrum using thin metallic meshes. The irregular polygonal metallic meshes with optimized dimensional parameters were fabricated on sapphire substrate, and exhibits efficient EMI SE of more than 20 dB in the wide frequency range of 1.7–18 GHz while maintaining high optical transparency with infrared rays. Additionally, the shielding effectiveness shows no obvious decline in high frequency range, exhibiting a balanced shielding property over a wide frequency spectrum. The results also indicate that mesh thickness is a critical parameter which make an impact on EMI shielding property over wide frequencies. This work is of great importance to practical applications in the protection of infrared optoelectronic systems from electromagnetic interference.

**2. Materials and Methods**

Figure 1(a) schematically illustrates the structure of the transparent EMI shielding window. It is composed of a sapphire substrate and two layers of metal mesh films on its surface. Sapphire is adopted in this study due to its outstanding physical features, including high infrared transparency, good thermal conductivity, high hardness, and ability to withstand high temperatures. The thickness of Au mesh layer is designed between 200 and 300 nm, which is sufficient to offer good conductivity for achieving effective microwave reflection. The Cr mesh is used as the adhering layer, and its thickness is much lower than that of the Au mesh. Besides the mesh-based shielding films, a layer of $Al_2O_3$ is employed as the covering layer, which can protect the mesh film from the erosion in harsh environment.

The mesh pattern design affects the optical properties of the transparent shielding material. Due to the centralized diffraction distribution, conventional meshes (like square metal grid) offer poor imaging quality [25]. To diminish the concentration of high-order diffraction and to minimize the impact of imaging caused by mesh shielding layer, the metal mesh is specifically designed to be with irregular patterns. The irregular mesh patterns are presented in Figure 1(b). Since the IR transmittance of the metal mesh is affected by the obscuration ratio, large equivalent period combined with narrow linewidth can achieve high IR transmittance. In this study, the equivalent period and linewidth of mesh patterns are designed to be 110 μm and 2.5 μm, respectively.

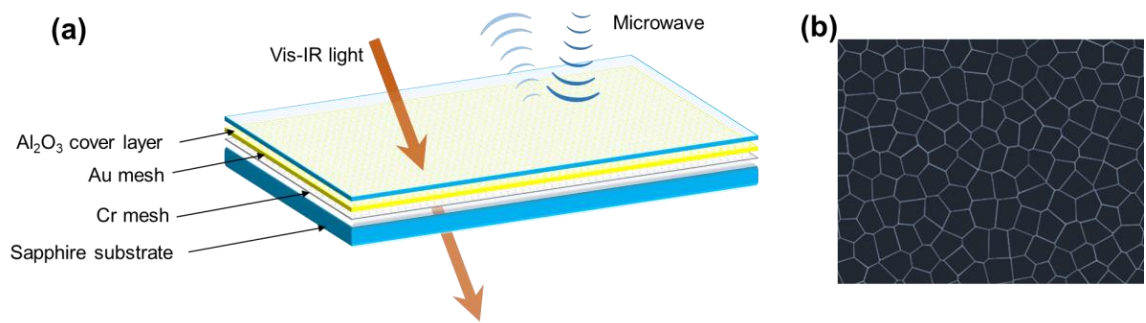

Figure 1. Design of IR transparent EMI shielding window. (a) Schematical illustration of the structure design and function of IR transparency and microwave reflection. (b) Mesh pattern of irregular polygons.

Metallic mesh films can be fabricated using various methods, including photolithography [23,26], nano-imprinting [27], and laser direct-writing technology [28]. To obtain meshes with fine lines, mature processes of ultraviolet (UV) photolithography was adopted in this study to prepare the samples. The specific fabricating process is illustrated in Figure 2(a). First, positive photoresist (AZ 5214E) was spin-coated onto the cleaned sapphire substrate. Then photoresist film (~2 μm in thickness) with mesh pattern obtained after exposure and development process. Afterwards, metal films (Cr/Au) were deposited by a magnetron sputtering system (PVD75, Kurt J. Lesker) in the presence of argon with a base pressure of $3.0 \times 10^{-6}$ Torr. The deposition rate of Cr and Au were 0.66 Å/s and 1.00 Å/s, respectively. The thin Cr metal layer is provided for better adhesion of the grid layer to the sapphire substrate. The photoresist was then dissolved to reveal the metal meshes in accordance with the designed mesh pattern. Figure 2(b) shows the micrograph of metallic meshes under magnification of 50X. It can be observed that the metallic wires are continuous and have no fractures, which ensures that the mesh coating has good electrical conductivity. According to Figure 2(c), the average linewidth of the mesh is measured to be ~2.5 μm.

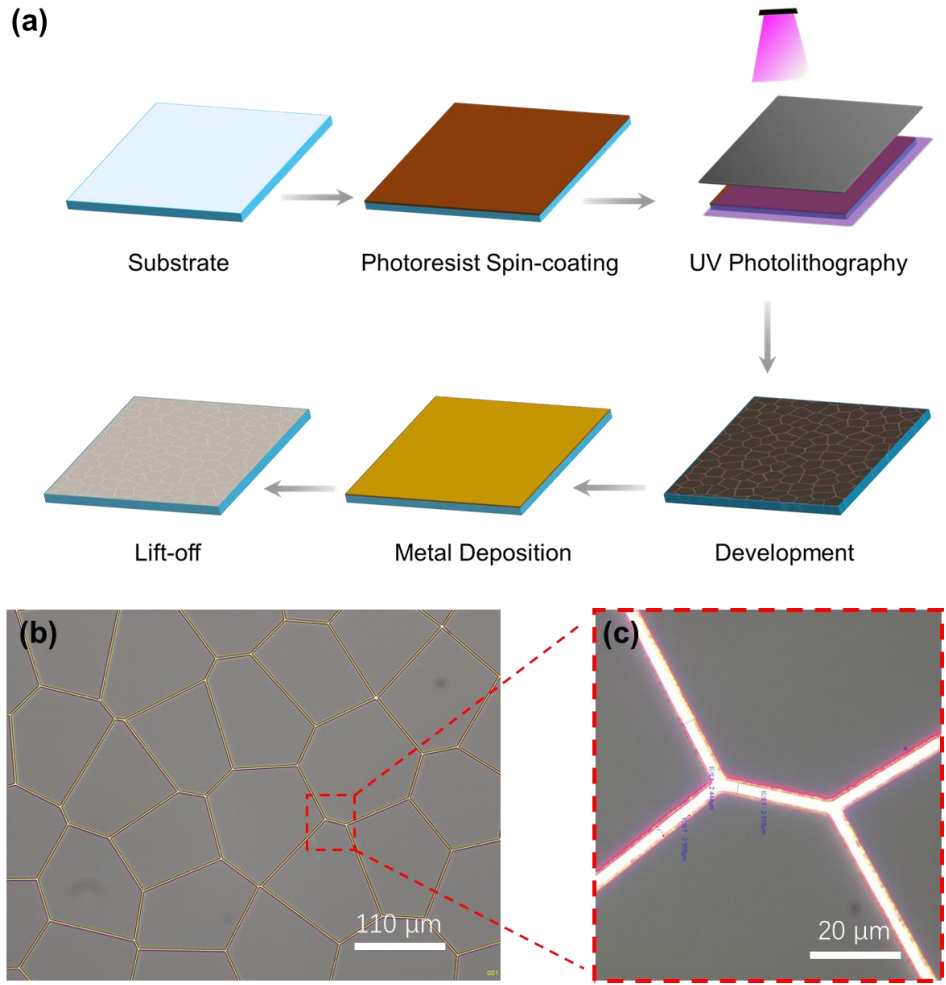

Figure 2. (a) Fabricating process of the IR transparent EMI shielding window. (b) Micrograph of metallic meshes under magnification of 50X. (c) Measurement of linewidth under magnification of 400X.

## 3. Results and Discussion

The optical transmittance of the sample in mid-wave IR region (3–5 μm) was measured using a Fourier transform infrared spectrometer (FTIR-650S, Gangdong Sci & Tech) under ambient conditions. The IR transmittance of the bare sapphire window, sapphire window with mesh coatings, and sapphire window with mesh coatings and cover layer are plotted in Figure 3. It can be seen that the bare sapphire window itself is highly transparent to IR light, exhibiting a high transmittance of 88.89% at 3.210 μm. Since the mesh has large effective apertures for light, the infrared transmittance of the window decreased by only 4.14% after the mesh coating was fabricated.

The most interesting aspect of the result is that the transmittance of the electromagnetic shielding window with cover layer is slightly increased by 1.42% and reaches 86.17%, which can be attributed to the destructive interference [29]. As a result, the electromagnetic shielding layer reduces the window's infrared transmittance only by 2.72%. With such a high transmission rate, the employment of the electromagnetic shielding window does not affect the detecting and imaging functions of optoelectronic system. It is worth noting that the IR transmittance starts to decrease at wavelengths greater than 4 μm and drops to ~48% at 5 μm. This is due to the increased IR absorption of $Al_2O_3$ in this region. In addition, it is worth mentioning that the transmission short-wave cutoff limit is about 0.19 μm [30], the EMI shielding window possesses high visible light transmittance as well.

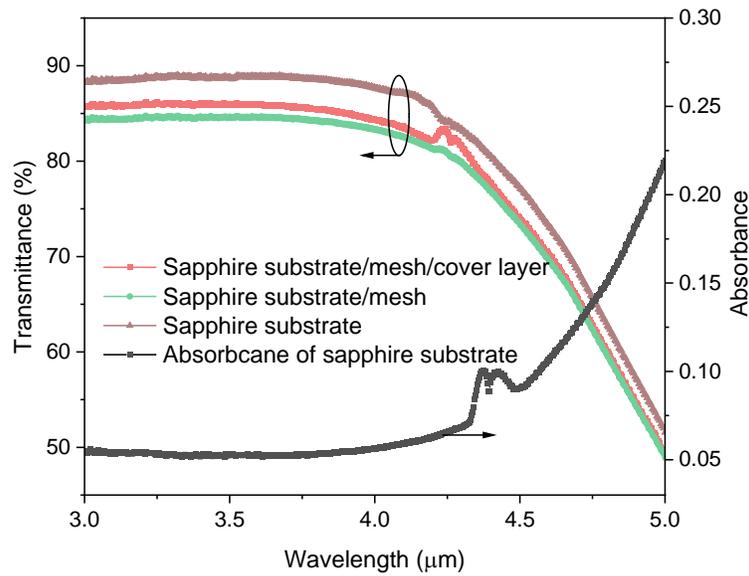

Figure 3. Optical properties of the transparent EMI shielding window.

Attenuation of electromagnetic waves by shielding materials is quantified by shielding effectiveness (SE), which is the logarithmic ratio of the incident power ($P_i$) to the transmitted power ($P_t$) [10]. Measurements of EMI shielding performance in this study were conducted over a wide frequency range of 1.7–18 GHz using the waveguide method, which is schematically illustrated in Figure 4(a). The EMI shielding measurement set-up mainly consists of a vector network analyzer (VNA, AV3672B, Ceyear) and waveguides, as is shown in Figure 4(b). The system was calibrated using the Transmission-Reflection-Load (TRL) technique before
7

testing. As-prepared samples were firmly clamped between the commissure of waveguides, and different sets of waveguides operating in L-, S-, C-, X-, and Ku-band were utilized to achieve shielding performance across a broad frequency spectrum. EMI SE values can be obtained through *S*-parameters collected by the VNA. Since the ratio of $P_i$ and $P_t$ is the measure of electromagnetic transmittance, the total SE can be expressed as [10,28]

$$\mathrm{SE} = -10\lg(P_t/P_i) = -10\lg(|S_{21}|^2) \tag{1}$$

According to transmission line theory, EMI Shielding effectiveness of a conductive film is mainly dependent on the sheet resistance [22,31], and the metallic mesh performs as a continuous conductive film at low frequencies[32]. The sheet resistance of a mesh film can be controlled by the structural parameters of mesh thickness, period, and linewidth. In this study, linewidth of the metal mesh is of 2.5 μm and the mesh period is 110 μm. The thickness of Cr and Au mesh layers are 35 nm and 285 nm, respectively. Figure 4(c) presents the measured EMI shielding performance of the mesh-coated sapphire window. The average EMI SE in the wide frequency range of 1.7–18 GHz is measured to be 23.5 dB, which can block more than 99% electromagnetic power and is sufficient for practical applications [33].

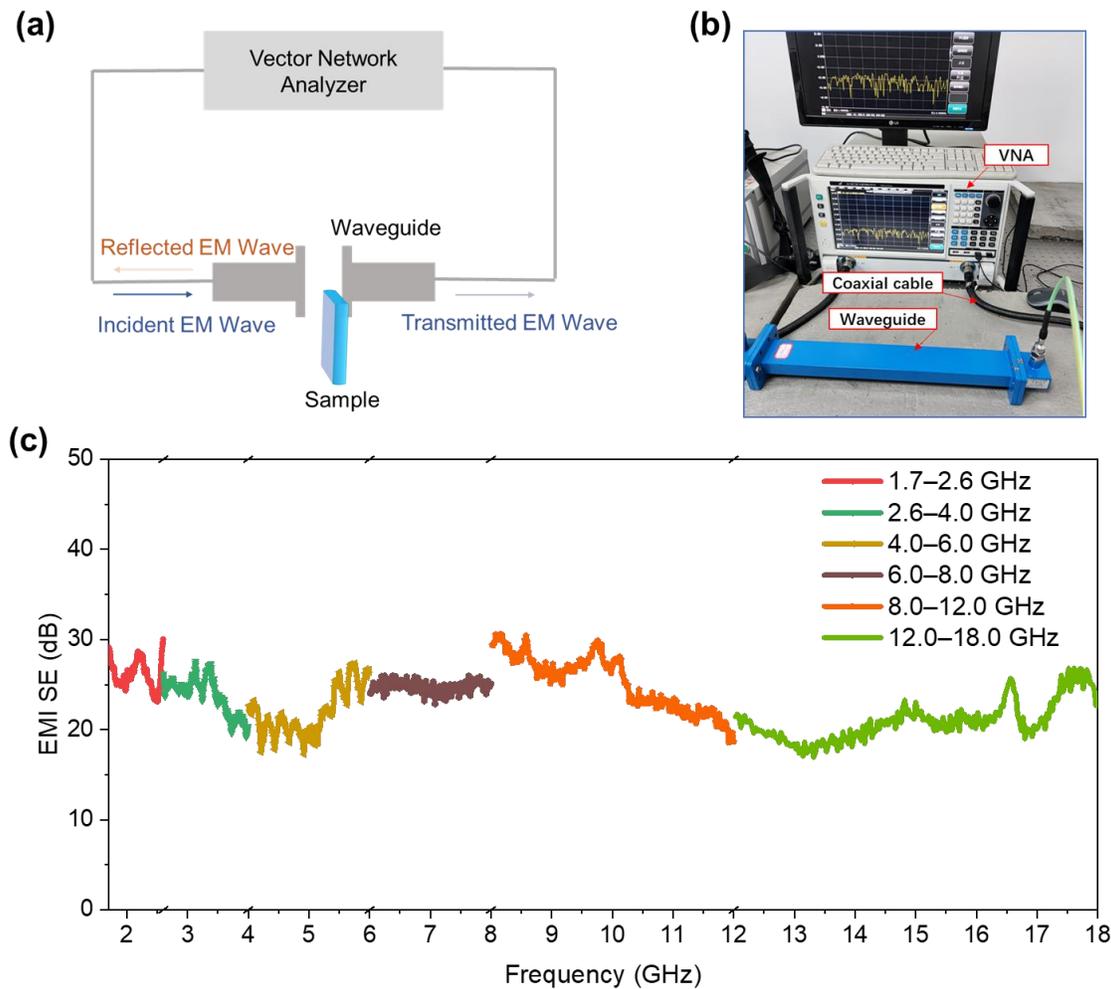

Figure 4. EMI shielding performance. (a) Schematic illustration of EMI SE measuring method based on waveguides. (b) The actual EMI SE measurement set-up. (c) Measured EMI SE of transparent electromagnetic shielding window over wide frequency range of 1.7–18 GHz

It has been reported that mesh film exhibits obvious EMI shielding degradation as higher frequencies. What is different in this study is that the shielding performance of mesh coating is balanced over the wide frequency range. For mesh coatings, the underlying cause of frequency-dependent property of EMI shielding can be explained by analyzing the transmission ($T$) of electromagnetic waves, which can be express as following in the case at normal incidence in free space [21]

$$T = \frac{4y^2}{1+4y^2} \quad (2)$$

where $y$ is the normalized admittance of the individual mesh film without a substrate. When the linewidth ($w$) of the mesh is larger than its thickness and much smaller than the period ($p$), the normalized admittance can be expressed as

$$y = -\frac{p}{\lambda}\left[\ln\left(\sin\frac{\pi w}{2p}\right)\right] \quad (3)$$

where $\lambda$ is the wavelength of the incident electromagnetic wave. According to equation (3), as the frequency gets higher and the wavelength gets smaller, the normalized admittance of the conductive mesh decreases. Therefore, the metal mesh exhibits a decrease in shielding efficiency at higher frequencies. In addition to the influence of the aperture structure on the transmission coefficient of the mesh, the variation of the sheet resistance of the mesh with frequency is also an important reason for the frequency dependence of its shielding effectiveness. The frequency dependance of sheet resistance for mesh coatings can be described as follows [34]

$$R_{mesh} = \frac{1}{\sigma\delta(1-e^{-t/\delta})}\frac{p}{w} \quad (4)$$

Here $\delta$ is the skin depth, and $\delta = \sqrt{1/(\pi f \mu \sigma)}$, $\sigma$ is the conductivity of the metal, $t$, $p$, and $w$ are the thickness, period, and linewidth of the mesh, respectively. As the frequency get higher, the sheet resistance of the mesh film increases.

Usually, the thickness of the metal mesh is designed to be greater than the skin depth in order to obtain higher shielding effectiveness. However, increasing the thickness only improve shielding performance at low frequencies. For the mesh coating with thickness much less than the skin depth, the SE is dominated by the reflection of incident plane wave at the air-mesh interface. When the skin depth decreases, the shielding effectiveness also decreases accordingly [28]. However, the skin depth of a conductive film decreases as the frequency gets higher.

A thicker metal mesh can reduce the sheet resistance of the mesh coating and enhance the EMI shielding performance at lower frequencies [34]. However, it contributes a little to the improvement of EMI SE at higher frequencies. Comprehensive consideration of the effects of mesh aperture structure, surface resistance and skin depth on shielding effectiveness, it is possible to alleviate the frequency dependence of shielding effectiveness by designing the proper mesh thickness for achieving a balanced shielding effectiveness in a wide frequency band. In this study, the mesh thickness is several hundred nanometers, much thinner than the skin depth of the metallic film. According to the measured result, average EMI SE of L- to Ku-band are 25.5 dB, 23.1 dB, 23.6 dB, 24.4 dB, and 20.9 dB, respectively. It can be seen that EMI SE is only weakly dependent on frequency, demonstrating that the thin metal film is beneficial to alleviate the dependence of EMI SE on frequency. Besides, the irregular mesh pattern design also contributes to the stable shielding property, which has been experimentally verified by Han et al. [24]. The thin meal mesh with irregular patterns in this study not only ensures the effective EMI shielding performance, but also reduces the frequency dependence of the shielding effectiveness, which is beneficial to obtain a balanced shielding effectiveness in a wide frequency band. The advantage of adopting the thin mesh is that a more stable shielding performance can be achieved over a wide frequency band. In addition, the design of the thin mesh facilitates costly and time-consuming mesh fabricating process, making it possible to achieve the desired shielding performance in an efficient manner.

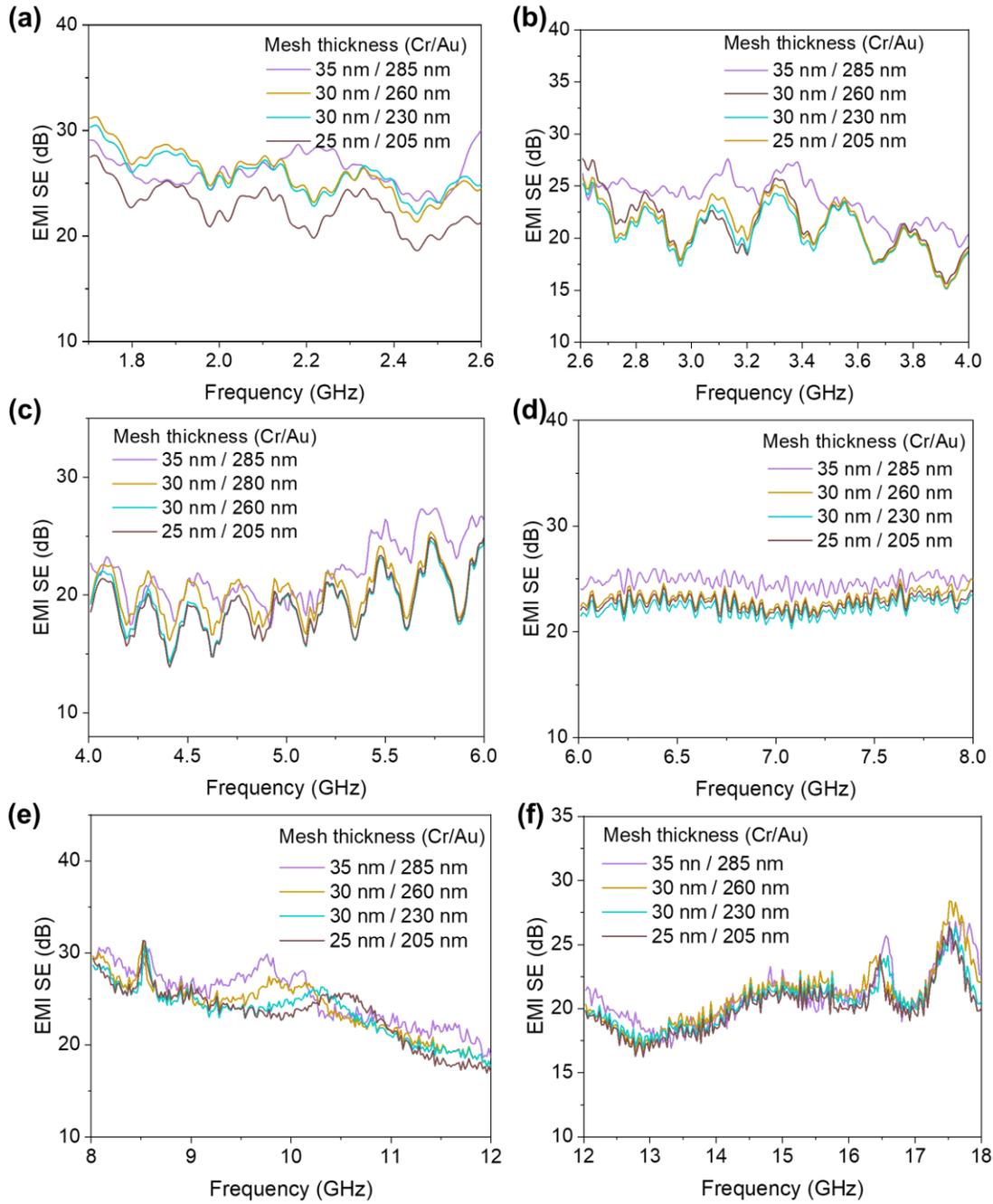

Figure 5. EMI shielding performance of metallic meshes with different thickness in varying frequency band. (a) 1.7–2.6 GHz. (b) 2.6–4 GHz. (c) 4–6 GHz. (d) 6–8 GHz. (e) 8–12 GHz. (f) 12–18 GHz.

Table 1. Summary of average SE of meshes with varying thickness in each frequency band

| Mesh thickness of Cr/Au | Average SE (dB) | | | | | |
| --- | --- | --- | --- | --- | --- | --- |
| | 1.7–2.6 (GHz) | 2.6–4 (GHz) | 4–6 (GHz) | 6–8 (GHz) | 8–12 (GHz) | 12–18 (GHz) |
| 35 nm/285 nm | 26.1 | 23.7 | 21.9 | 24.7 | 25.2 | 20.9 |
| 30 nm/260 nm | 25.9 | 21.3 | 20.3 | 23.1 | 23.9 | 20.9 |
| 30 nm/230 nm | 25.7 | 20.6 | 19.2 | 22.2 | 23.5 | 20.4 |
| 25 nm/205 nm | 22.7 | 21.1 | 19.1 | 22.7 | 23.3 | 20.0 |

Figure 3a-f present the EMI shielding performance of metal meshes with different thickness in six measured frequency bands. One can see that the thickness variation has different effects on the shielding effectiveness in different frequency bands. Close inspect of the figures show that the effect of thickness variation on shielding effectiveness is more evident at lower frequencies. In the frequency range of 1.7–2.6 GHz, the average EMI SE values of samples with thickness of 320 nm and 230 nm are 26.1 dB and 22.7 dB, respectively (Figure 5(a)). According to Figure 5(f), however, the average EMI SE of the two samples are 20.9 dB and 20.0 dB, showing that the effect of thickness variation on shielding effectiveness gradually weakens as the frequency increases. Specific EMI SE values of the samples in different frequency bands are obtained and summarized in Table 1. As the frequency gets higher, the difference in shielding effectiveness exhibited by mesh coatings with different thicknesses gradually decreases, indicating that the variation of mesh thickness has little effect on the shielding effectiveness at high frequencies.

## 4. Conclusions

In summary, a metallic mesh-based transparent EMI shielding window featuring high IR transparency as well as balanced electromagnetic shielding performance over wide shielding frequency spectrum has been demonstrated. The mesh coating is fabricated on sapphire window by UV photolithography, providing both high IR transmittance of 85.1% and favorable EMI SE of 23.5 dB. Meanwhile, the shielding effectiveness is balanced in the wide frequency band and only weakly dependent on frequency. Moreover, optimization of EMI shielding performance by varying the thickness of the mesh film has also been verified. The extra benefit of this design is that the process of metal deposition, which used to be time-consuming and energy-intensive during thick mesh preparation, is considerably decreased. This facilitates the fabrication of efficient transparent electromagnetic shielding materials with balanced wideband shielding performance. All the outstanding properties make the mesh-coated sapphire window an ideal EMI shielding component for optoelectronics systems.

**Author Contributions:** Conceptualization, X.H.; data curation, Y.L. and W.L.; formal analysis, Y.L. and X.H.; investigation, Y.L; methodology, J.P.; validation, K.W., L.Y. and J.P.; supervision and project administration, X.H. and P.L.; writing – original draft, Y.L.; writing – review and editing, K.W., L.Y., and X.H. All authors have read and agreed to the published version of the manuscript.

**Funding:** This research was funded by the National Natural Science Foundation of China (No.61801490).

**Conflicts of Interest:** The authors declare no conflicts of interest.

12. Han, J.; Wang, X.; Qiu, Y.; Zhu, J.; Hu, P. Infrared-transparent films based on conductive graphene network fabrics for electromagnetic shielding. *Carbon* **2015**, *87*, 206-214, doi:10.1016/j.carbon.2015.01.057.
13. Chen, W.; Liu, L.-X.; Zhang, H.-B.; Yu, Z.-Z. Flexible, Transparent, and Conductive Ti3C2Tx MXene-Silver Nanowire Films with Smart Acoustic Sensitivity for High-Performance Electromagnetic Interference Shielding. *Acs Nano* **2020**, *14*, 16643-16653, doi:10.1021/acsnano.0c01635.
14. Van Viet, T.; Duc Dung, N.; Nguyen, A.T.; Hofmann, M.; Hsieh, Y.-P.; Kan, H.-C.; Hsu, C.-C. Electromagnetic Interference Shielding by Transparent Graphene/ Nickel Mesh Films. *Acs Applied Nano Materials* **2020**, *3*, 7474-7481, doi:10.1021/acsanm.0c01076.
15. Zhang, C.; Ji, C.; Park, Y.B.; Guo, L.J.J.A.O.M. Thin-Metal-Film-Based Transparent Conductors: Material Preparation, Optical Design, and Device Applications. *Advanced Optical Materials* **2021**, *9*, 2001298.
16. Wan, Y.J.; Zhu, P.L.; Yu, S.H.; Sun, R.; Wong, C.P.; Liao, W.H.J.S. Anticorrosive, ultralight, and flexible carbon-wrapped metallic nanowire hybrid sponges for highly efficient electromagnetic interference shielding. *Small* **2018**, *14*, 1800534.
17. Wang, Y.-Y.; Zhou, Z.-H.; Zhou, C.-G.; Sun, W.-J.; Gao, J.-F.; Dai, K.; Yan, D.-X.; Li, Z.-M.J.A.a.m.; interfaces. Lightweight and robust carbon nanotube/polyimide foam for efficient and heat-resistant electromagnetic interference shielding and microwave absorption. *ACS Appl. Mater. Interfaces* **2020**, *12*, 8704-8712.
18. Zhu, X.; Xu, J.; Qin, F.; Yan, Z.; Guo, A.; Kan, C. Highly efficient and stable transparent electromagnetic interference shielding films based on silver nanowires. *Nanoscale* **2020**, *12*, 14589-14597, doi:10.1039/d0nr03790g.
19. Yuan, C.; Huang, J.; Dong, Y.; Huang, X.; Lu, Y.; Li, J.; Tian, T.; Liu, W.; Song, W. Record-High Transparent Electromagnetic Interference Shielding Achieved by Simultaneous Microwave Fabry-Perot Interference and Optical Antireflection. *Acs Applied Materials & Interfaces* **2020**, *12*, 26659-26669, doi:10.1021/acsami.0c05334.
20. Han, Y.; Lin, J.; Liu, Y.; Fu, H.; Ma, Y.; Jin, P.; Tan, J.J.S.r. Crackle template based metallic mesh with highly homogeneous light transmission for high-performance transparent EMI shielding. *Sci. Rep.* **2016**, *6*, 1-11.
21. Tung, P.D.; Jung, C.W. High Optical Visibility and Shielding Effectiveness Metal Mesh Film for Microwave Oven Application. *Ieee Transactions on Electromagnetic Compatibility* **2020**, *62*, 1076-1081, doi:10.1109/temc.2019.2927923.
22. Voronin, A.S.; Fadeev, Y.V.; Govorun, I.V.; Podshivalov, I.V.; Simunin, M.M.; Tambasov, I.A.; Karpova, D.V.; Smolyarova, T.E.; Lukyanenko, A.V.; Karacharov, A.A.; et al. Cu-Ag and Ni-Ag meshes based on cracked template as efficient transparent electromagnetic shielding coating with excellent mechanical
16